# Relative Mobility of Human Transferrin Domains Accounts for Its Efficient Recognition and Recycling

Haleh Abdizadeh, Ali Rana Atilgan, Canan Atilgan*

Sabanci University, Faculty of Engineering and Natural Sciences,

Tuzla, 34956 Istanbul, Turkey

*Corresponding author

Faculty of Engineering and Natural Sciences
Sabanci University, Tuzla 34956, Istanbul, Turkey
*e-mail*: canan@sabanciuniv.edu(CA)
*telephone*: +90 (216) 483 9523
*telefax*: +90 (216) 483 9550

# Abstract

Human serum transferrin (hTf) transports ferric ions in the blood stream and inflamed mucosal surfaces with high affinity and delivers them to cells via receptor mediated endocytosis. A typical hTf is folded into two homologous lobes; each lobe is further divided into two similar sized domains. Three different crystal structures of hTf delineate large conformational changes involved in iron binding/dissociation. However, whether the release process follows the same trend at serum (~7.4) and endosomal (~5.6) pH remains unanswered. The specialized role of the two lobes and if communication between them leads to efficient and controlled release is also debated. Here, we study the dynamics of the full structure as well as the separate lobes in different closed, partially open, and open conformations under the nearly neutral pH conditions in the blood serum and the more acidic one in the endosome. The results corroborate experimental observations and underscore the distinguishing effect of pH on the dynamics of hTf. Furthermore, in a total of 2 μs molecular dynamics simulation of different forms of hTf, residue fluctuations elucidate the cross talk between the two lobes correlated by the peptide linker bridging the two lobes at serum pH, while their correlated motions is lost under endosomal conditions. At serum pH, the presence of even a single iron on either lobe leads C-lobe fluctuations to subside, making it the target for recognition by human cells or hostile bacteria seeking iron sequestration. The N-lobe, on the other hand, has a propensity to open, making iron readily available when needed at regions of serum pH. At endosomal pH, both lobes readily open, making irons available for delivery. The interplay between the relative mobility of the lobes renders efficient mechanism for the recognition and release of hTf at the cell surface, and therefore its recycling in the organism.

# Author Summary

Human serum transferrin (hTf), a protein with bilobal structure, is involved in iron trafficking between the sites of iron storage and intake. Three different conformational states of hTf operate based on many environmental constraints such as pH, chelators, temperature, salt concentration and lobe-lobe interactions. Lobe-lobe interactions and the ability to capture iron by each lobe have been suggested to play major roles in the dynamics of interconversion between conformational substates. We distinguish the dynamical variety of different states of hTf by devising extensive all-atom MD simulations. We show that a starting conformation may be destabilized by manipulating protonation states of some distinct residues. At neutral pH, we observe major structural changes including an advance toward an open N-lobe from the initial steps of the simulation, the whole structure advancing to a partially open structure, also captured by x-ray crystallography. We explain the transferrin cycle, i.e. the recognition of hTf at the cell surface in iron loaded form and its release by the cell after extraction of iron, by a simple model that takes into account the changes in relative mobility of the two iron-carrier lobes in the presence/absence of iron.

# Introduction

All mammals require iron in fundamental metabolic processes to fulfill the key roles of enzymatic electron transfer, proliferation of iron-sulfur clusters and oxygen delivery [1]. Cellular iron deficiency prevents cell growth and leads to cell death. However, excess iron engages in Fenton reaction and facilitates formation of destructive hydroxyl radicals that attack lipids, proteins and nucleic acids. Hence, iron offers a biologically vital, but potentially dangerous role in the human body. To control iron levels, sophisticated mechanisms have evolved to acquire, store and deliver iron [2-5]. One mechanism of iron circulation in the body is via the transferrin (Tf) family of proteins found in multicellular organisms from insects to humans. This is a group of nonheme iron-binding glycoproteins with a single peptide chain of nearly 700 amino acids playing the role of $Fe^{+3}$ transport from the sites of intake to the circulatory system, cells and tissues [6]. Three main proteins of this group have been identified as serum transferrin (sTf), lactoferrin (Lf) and ovotransferrin (OTf). Lf and OTf function as antimicrobial agents and are present in secretory fluids such as milk or tears, and in white blood cells [7-9]. Any Tf family member consists of two lobes designated as the N-lobe and the C-lobe, which are further divided into two subdomains termed $N_1/N_2$ and $C_1/C_2$. There is strong homology (~60%), both in sequence and tertiary structure, between the members, and between the two lobes of any given Tf [7].

X-ray crystal structures of transferrin family members reveal that human serum transferrin (hTf) exhibits three different conformations, the iron-free fully open apo conformation, the fully closed iron loaded holo conformation and iron loaded partially open conformation exhibiting a structure intermediate to that of the apo and holo hTf [6,7,10]. The conformational difference between A and H forms involves the opening at the two hinges between the $N_1/N_2$- and $C_1$-/$C_2$- subdomains, and iron release making use of their twist and bending [7] [11]. The C-lobe in PO has the same fold as H. However, the N-lobe adopts a unique structure with a partially open cleft [6].

In most cell types, iron intake is initialized through a process known as the transferrin cycle. The ligand-receptor interaction between hTf/transferrin receptor (TfR) complex induces clathrin-mediated endocytosis [12]. The lower pH of the endosome leads to conformational shift in both hTf and TfR and promotes iron release. Subsequently, released iron is reduced to $Fe^{+2}$ by members of the STEAP family of metalloreductases. Meanwhile, apo-hTf and TfR are recycled to the cell surface for reuse [2,3]. The tertiary structure of hTf is crucial for TfR recognition. The higher affinity of TfR towards holo-hTf in the extracellular region (pH ~7.4) leads to localization of hTf into the endosome. In the endosome, TfR remains bound to apo-hTf, resulting in the trafficking of apo-hTf from the endosome to extracellular membrane where hTf is finally released (see figure 1 in ref. [12]).

The primary question concerning hTf is the mechanism of iron binding with nanomolar affinity ($K_d$~$10^{-22}$ M) and yet its prompt release at the reduced pH of endosome. Specifically, the conformational changes prior to and during the iron release process are not fully understood despite considerable efforts devoted to study the molecular structure and kinetics of hTf. Furthermore, understanding the molecular basis of the N- and C-lobe characteristics and major differences between their dynamics may help develop a model for the iron release mechanism. Computational approaches to study hTf dynamics to date have only analyzed the conformational details of the N-domain of hTf [11,13]; to the best of our knowledge, this is the first computational study of the full structure hTf, which also addresses the crosstalk between its two domains.

It is hypothesized that hTf has specific features which differ at physiological and endosomal pH (see e.g. [12]). The structural difference between the two lobes of hTf account for the nuances in their biological functionalities as well as their thermodynamic and kinetic properties in different environments [2]. Dissimilarities arise in part because of second shell residues. These do not contribute to iron binding, but participate in a complex hydrogen bonding network with iron coordinating residues. In the N-lobe of hTf, second shell residues are comprised of K206 and K296, forming the so-called "dilysine trigger," positioned in $N_2$ and $N_1$ subdomains, respectively [14]. In the C-lobe, the dilysine trigger is replaced by a triad of residues consisting of K534 in the $C_2$-subdomain, R632 and D634 in the $C_1$-subdomain [15]. It has been suggested that the hydrogen bond network created by the second shell residues at physiological pH is disrupted upon protonation processes occurring at the endosomal pH, assisting domain opening [14]. In a computational study, open and closed forms of iron free N-lobe of hTf was investigated theoretically [11]. $pK_a$ calculations showed that the low $pK_a$ of K206 (~7.4) dictates deprotonated state of this residue at pH 7.4, while K296 is protonated. The authors tested the postulation that further protonation of K206 at the reduced pH of the endosome would result in a repulsion between the two lysines and would assist lobe opening for iron release. Based on 1.5 ns MD simulations produced under the CHARMM force field, they showed that the repulsion between the two lysines may be locally accommodated without the need for domain opening. Therefore, an alternative mechanism was proposed therein, whereby K206/K296 may facilitate the protonation of Y188 followed by domain opening. In addition, MD simulations and molecular mechanics-Poisson-Boltzmann surface area free energy calculations (MM-PBSA) [16] relate the enhanced mobility of the open form compared to the closed form of apo hTf to entropic effects. However, the aforementioned simulations do not model iron and carbonate which are expected to strongly affect hTf properties. In addition, the possible effect of the C-lobe on the N-lobe dynamics (or the reverse) is not considered, since that work relies on the MD simulations of the N-lobe only.

In another study, Mujika et al. [13] performed 60 ns MD simulation under the GROMACS 4.5.3 force field to examine different protonation states of iron and aluminum loaded N-lobe of hTf. They observe that protonation of Y188 is indeed a necessary step in domain opening while protonation state of the dilysine trigger alone does not lead to this conformational change. Therein, it was also confirmed that hinge-bending and hinge-twist motions precede metal release. The authors further carried out hybrid quantum mechanics/molecular mechanics simulations on the N-lobe to analyze the dynamics of the metal loaded hTf in the blood and cellular pH [17]. These analyses show that at serum pH the iron coordination is octahedral, while it is distorted at endosomal pH, once H249 leaves the first coordination shell and locates to the second coordination sphere. Therefore, neutralization of the imidazolate group of H249 seems to be a necessary step in the iron release process. Thus, independent of the force field type and simulation time length, previous studies imply that the protonation state of Y188 is the key event in iron release from the N-lobe regardless of the second shell residue status.

Motivated by our earlier work where conformational changes were related to pH changes [18-20], here we systematically study the dynamics of full length hTf. Previously, we have carried our extensive MD simulations to investigate the conformational multiplicity and dynamics of calcium loaded calmodulin (CaM) at different protonation states and using different initial structures [18,20]. We concluded that manipulating the pH sensitive residues of CaM prompts a conformational switch from extended structures to more compact ones. The equilibrated structures also corroborated observations from FRET and NMR experiments. Therefore, we proposed that slow events might be triggered by altering the charge distribution of the protein. Interestingly, these acidic residues with upshifted pK$_a$s were found to be located in allosterically significant regions. In another study, ferric binding protein (FBP) in holo and apo forms were investigated using Perturbation Response Scanning method [21]. A surface residue that is mechanically coupled to the iron binding site was proposed to play a significant role in manipulating the conformational change. Following detailed MD simulations on apo FBP disclosed the allosteric effect of this residue on the binding/release dynamics of the synergistic anion phosphate [19].

This is the first MD study of the full crystal structure of the hTf protein, which reports its behavior under various initial environmental conditions in the presence of $Fe^{+3}$ and the synergistic carbonate. Our aim is to examine conformational dynamics of these forms and elucidate molecular functioning mechanisms of hTf. We communicate results of a series of extensive MD simulations (of minimum duration of 60 ns) performed on the three resolved conformations (H, A and PO) of the protein; we also manipulate the pH sensitive residues. To examine the routes that may be accounted for the control of coordination between the lobes, we have studied monoferric full length hTf, as well as the separate N- and C-domains.

Analysis of the resulting trajectories offers details on the conformational shift and iron binding mechanism. We analyze the structural dynamics through identifiers based on reduced degrees of freedom defined specifically for hTf, defining both the opening modes of motion within the separate lobes and the relative positioning of the lobes. Key events leading to, or preventing, conformational change are discussed. We monitor the path sampled between different conformational states identified by MD simulations and we evaluate the effect of changing the charge states of key residues. The molecular mechanisms that lead to the observed effects, their relationship to the experimental data, and the consequences of the observations that enhance our understanding of the dynamics and function of hTf are outlined.

# Materials and Methods

**Human serum transferrin**. The overall structure of bilobal hTf exhibits two homologous halves termed N-lobe (residues 1-331) and C-lobe (339-679). It is further divided into $N_1$-(1-95 and 247-331), $N_2$-(96-246), $C_1$-(339-425 and 573-679) and $C_2$-subdomains (426-572). Intra/inter-lobe interactions and an unstructured peptide linker (332-338) stabilize the tertiary structure of the protein. The linker provides hTf an overall flexibility [6]. The three dimensional structure of holo hTf is schematically shown in figure 1a, and displayed in figure S1.

The experimentally determined holo and partially open forms are fully loaded with two $Fe^{+3}$ ions located in a cleft formed between the subdomains. The coordination of the two ferric ions in holo hTf is shown in figure 1b. Highly conserved iron liganding residues in N- (C-)lobes are comprised of the phenolate oxygen atom of a tyrosine in the hinge at the edge of $N_2$ ($C_2$), a second tyrosine within $N_2$ ($C_2$), carboxylate oxygen atom of an aspartic acid, the sole ligand from $N_1$ ($C_1$) and the imidazole nitrogen of a histidine at the hinge bordering $N_1$ ($C_1$) [5]. Two oxygen atoms from synergistic anion, carbonate, complete the distorted octahedral coordination of iron. Carbonate is anchored in place by conserved arginine and threonine residues. Its presence is crucial for high affinity iron binding [5,22].

The existing x-ray structures of $Fe^{+3}$-loaded hTf are either in closed or a partially open form. There are many examples of full structure and individual lobes of the closed form in the protein data bank (PDB) We select that with PDB code 3V83 whereby the 677 residue coordinates of the chain B of the crystal cell is utilized due to smaller number of missing atoms; we shall call this structure the holo form (H) in the rest of the manuscript [10]. The partially open (PO) form is represented by the 3QYT coded structure, and has 679 reported residues [6]. The open form has the 2HAV code and the first three residues are not present in the crystal structure [7]; we shall refer to this structure as the apo form (A) throughout. These structures have been determined at 2.1, 2.8 and 2.7 Å resolution, and were crystallized at pH of 7.5, 7.4 and 7.0, respectively. We report the RMSD comparison for the overall structure as well as the N- and C-lobes of various x-ray structures in Table 1. In particular, we find that the C-lobe of the PO state is similar to that of H (0.5 Å RMSD), while its N-lobe is similar to that of A (1.7 Å RMSD).

We have previously introduced reduced degrees of freedom to successfully distinguish the conformational changes in CaM [20]. We use the same efficient approach here to distinguish between the different conformations of hTf. For hTf, we project the 3$N$-dimensional conformational space into three visually tractable ones. These are bending angles within the N- and the C-lobes ($\theta_N$ and $\theta_C$) and a torsional angle defining the relative positioning of the two lobes ($\varphi$). The former is defined as the bending angle between the center of mass of each subdomain in every lobe and a hinge residue; we choose $C_\alpha$ atom of Y95 and Y426 as the hinge residues in the respective lobes, which also bind iron. The torsional angle is defined by four points: the hinge residues in the N-lobe ($C_\alpha$ atom of Y95), two end points of the peptide linker ($C_\alpha$ atoms of residues 331 and 339), and the hinge residue in the C-lobe ($C_\alpha$ atom of Y426). These points and the reduced degrees of freedom are schematically shown in figure 1a.

**Selection of protonation sites for mimicking environments at pH 7.4 and 5.6.** We mimic the different pH environments by assigning the most probable states of ionizable groups according to their p$K_a$ values which were calculated by PROPKA and H++ servers [23,24]. PROPKA relies on the accurate calculation of the shifts in free energies and it is based on an improved description of the desolvation and the dielectric response of the protein. The H++ server uses a different approach where the titration curves are directly represented as a weighted sum of Henderson-Hasselbalch curves of decoupled quasi-sites. We report values for the server settings of 150 mM salinity, external dielectric of 80 and internal dielectric of 10. We list the values in Supplementary Table S1 only for those residues where the calculated $pK_a$ is different from the standard value by more than 2 units. We also report calculated $pK_a$ values from ref. [11] for the N-lobe residues, based on the closed iron free structure of the N-lobe. These corroborate PROPKA and H++ server results for all cases except for two Tyr residues, whose role we will further discuss in this manuscript.

Of the acidic Asp and Glu residues, those with upshifted $pK_a$ values into the physiological range occur only in at most one of the two servers. Since the probability of an acidic residue being charged at 7.4 pH is greater than 0.9 for $pK_a < 6.6$, we assume that their most probable state carries negative charge at this pH as well as at endosomal pH. Similar argument applies to the basic Arg residues and all but one of the Lys; probability of being protonated at this pH is greater than 0.9 for $pK_a > 8.2$, so these are kept positively charged in all simulations. Finally, we observe downshifted values for K206 residues on both servers. We keep it positively charged since the calculations in ref. [11] also indicates it would be charged at the pH values we are interested in here. Note that, since the two Lys residues involved in the proposed dilysine trigger [14] (K206/296) are protonated in all our MD simulations, we will directly test if protonating them leads to the direct conformational switching of hTf.

All His residues have either the standard values, or are downshifted, having neutral side chains at pH 7.4. Thus, in the MD simulations of the systems at the higher pH, the standard protonation state is adopted for all His. To mimic the most probable states at endosomal pH, His residues with downshifted $pK_a$s (those listed in Table S1) are kept neutral while the rest of them are protonated.

Finally, Tyr residues directly coordinating the iron display downshifted values (Table S1). In particular, Y188 and Y517 exhibit a downshift in their values in all $pK_a$ predictions. We will change the protonation states of these residues in various combinations to test if a conformational switch may be triggered in the holo structure by the uptake/release of protons at these sites.

**Molecular dynamics simulations.** The NAMD package is used to model the dynamics of the protein-water systems [25]. Solvation is achieved via the VMD 1.9.1 program with solvate plug-in version 1.2 [26].k2 The protein is soaked in a cubic solvent box such that there is at least a 10 Å layer of solvent in each direction from any atom of the protein to the edge of the box. Ionic strength (IS) in all the simulations is 150 mM. Charge neutrality is obtained by adding an adequate number of $Na^+$ and $Cl^-$ ions. The CharmM22 all-atom force field is used to parameterize the topology of the protein atoms [27]. The force field parameters for synergistic carbonate ion and $Fe^{+3}$ reported in the literature are adopted [21,28]. TIP3P model is used to describe water molecules [29]. Long range electrostatic interactions are calculated by the particle mesh Ewald method [30], with a cutoff distance of 12 Å and a switching function at 10 Å. RATTLE algorithm [31] is applied to use a step size of 2 fs in the Verlet algorithm [30]. Temperature control is carried out by Langevin dynamics with a dampening coefficient of 5/ps. Pressure control is attained by a Langevin piston. The system is run in the NPT ensemble at 1 atm and 310 K until volumetric fluctuations are stabilized and they maintain the desired average pressure.

Ten sets of simulations were performed with various initial starting conditions and perturbation scenarios (Table 2). The different simulation conditions are labeled with the initial structure, H, A, or PO, for the full length holo, apo and PO structures, respectively; PDB codes used for each is listed in Table 1. The number of atoms in each system and equilibrated water box sizes are reported in Table 2. We have focused on five major issues: (i) The role of the protonation states of the Tyr residues in contact with the $Fe^{+3}$ ions in holo hTf (H, H* and H** systems); (ii) the similarity of the conformation triggered by neutralizing the aforementioned Tyr residues in the H state with the A and PO states (H, A and PO systems); (iii) the effect of having a single bound $Fe^{+3}$ ion on the H state (H, $H^{Fe+N}$ and $H^{Fe+C}$ systems); (iv) the role of the communication between the N- and C-lobes on the overall dynamics (H, $H_N$ and $H_C$ systems); and (v) the effect of lowering the pH (H and $H^\dagger$ systems). Each condition has at least 60 ns of sampling time, and at least 200 ns for those displaying non-equilibrated behavior by the end of 60 ns. For the H system, we have performed an additional, independent run of 150 ns to check the reproducibility of the results. We have also prolonged the original H simulation to 500 ns in case significant changes are observed in the relative local and global positioning of the two lobes, quantified by the RMSD of different regions (for details, see Results and Discussion).

## Result and Discussion

Many factors affect the distribution of conformational substates of hTf, including pH, chelators, temperature, salt concentration and lobe-lobe interactions, with current active research being focused on the effect of these factors on mutated and wild type hTf [12,32-37]. At pH 7.4, four major populations (namely, differric, two monoferric and apo hTf) have been suggested, based-on the iron occupancy of the deep cleft formed between the subdomains in each lobe of hTf [38]. Lobe-lobe interactions and the ability to capture iron by each lobe have been suggested to play a major role in the dynamics of interconversion between conformational substates [32]. Due to the difficulties in the experimental observation and quantification of hTf dynamics, all-atom MD simulations might help clarify some of these issues. However, the conformational switching at physiological pH is not expected to be observed on the time scales reachable by conventional simulations. Therefore, we seek local changes on the protein structure that will trigger the slow process. We therefore define perturbation scenarios under which the conformations sampled by hTf may be manipulated (Table 2).

**Domain opening is coupled to the protonation states of iron coordinating tyrosines.** In the holo form of hTf, it has been proposed that the iron coordinating Tyr residues have downshifted $pK_a$ values such that they are negatively charged when $Fe^{+3}$ is bound [11,39,40], also corroborated by various $pK_a$ calculations, particularly for Y95, Y188, and Y517 (Table S1). Therefore, we first run simulations to test that the systems are indeed stable when the protonation states of Tyr residues are changed (see figure S2 for the RMSD profiles over the course of 60 ns). We find that in H** system, where all the Tyr residues in the two $Fe^{+3}$ binding sites are deprotonated, the initial holo structure is stable with an overall, N-lobe, and C-lobe RMSD of 2.7±0.3, 2.1±0.1 and 2.4±0.3 Å, respectively, averaged over the last 50 ns. In the H* system, where Y188/Y517 are deprotonated and Y95/Y426 are protonated, we observe similar behavior to H** in both lobes. The system is stable around the initial conformation with an overall, N-lobe, and C-lobe RMSD of 2.8±0.3, 2.5±0.2, and 1.9±0.1 Å, respectively, averaged over the last 50 ns. The iron is still engaged in the binding region. As listed in Table 3, the liganding residue – $Fe^{+3}$ distance is within at most 2 Å of the initial, energy minimized structure in both the H* and H** simulations. This is contrary to the observations in the H run, which we discuss next by assessing how the systems respond to neutralizing iron coordinating Tyr residues in holo hTf.

While it is expected that the Tyr residues in the iron binding site would be neutral in the apo form, their ionization states in the holo form are debated and the $pK_a$ calculations also provide different results for these residues. E.g. PROPKA assigns $pK_a$ values greater than 12.0 to all these Tyr residues in the apo form, compared to 6.0 to Y188 and 4.1 to Y517 in holo form (the holo values are listed in Table S1). The RMSD profiles in a 200 ns run of the H system where all four iron coordinating Tyr residues are neutralized is displayed in figure 2a, and the first 60 ns of this run is also shown in figure S2 for comparison with the H** and H* runs. Large changes are observed in the whole protein almost instantly, leading to RMSD from the initial structure of up to 6.7±0.5 Å. In the first 60 ns, the curves indicate the N-lobe is more mobile than the C-lobe. In fact, it readily moves towards a more open conformer within the first 10 ns. Thus, the initial conformational change mostly involves motions in the N-lobe rather than global rearrangements, despite the fact that coordinating Tyr residues are neutralized in both lobes. After 100 ns, the C-lobe also displays larger deviations from the initial structure, moving towards a more open state. By the end of 200 ns, all four Tyr are ca. 6 Å away from the $Fe^{+3}$ compared to the ca. 2 Å value in the initial structure (Table 3). The Tyr residues in both lobes are the first ones to leave the binding region. The coordinating D63 and D392 in both lobes remain in contact with the iron, in conformity with mutagenesis studies which have shown that if Asp residue vacates the coordination shell, the iron will have enough freedom to leave the binding cleft [41]. In addition, the coordinating His on the C-lobe (H585) also remains in contact, while that on the N-lobe (H249) has moved away (Table 3). We conclude that the protonation states of the key Tyr residues are responsible for triggering the lobe opening, starting with the N-lobe.

**Partially open state is a stable intermediate of hTf.** To check if the new conformation attained by H is similar to other experimentally determined N- and/or C-lobe structures, we superimpose the trajectories of the lobes on the PO and apo form x-ray structures (supporting figure S3). We find that the N-lobe is in fact quickly driven towards a conformation similar to that in the apo form (2.9 Å RMSD) or the PO form (2.4 Å RMSD) by the triggering event. The C-lobe, on the other hand, is more similar to that in the holo form structure, despite a larger RMSD of 4.4 Å by the end of 200 ns. Overall, the final structure attained after the triggering event is more similar to the PO structure, with an open N-lobe and a closed C-lobe. We note that no additional events are observed when this trajectory is prolonged to 500 ns (data not shown). In addition, the initial triggering event on the N-lobe is observed within 20 ns in an independent, 150 ns long run carried out on the same system.

Figure 2b provides information on the dynamics of the PO system. It fluctuates around equilibrium up to 100-120 ns, then exhibits a small jump in RMSD. In the new equilibrated state, coordination of iron in the N-lobe binding cleft changes to what we observe by the end of 200 ns simulation of H. H249 and D63, which are 10.1 and 6.0 Å away from the $Fe^{+3}$ in the initial structure, respectively, move to a distance similar to that observed in the end of H simulations (Table 3). In accordance with this conformational change, the Y95 and Y188 move away to 6.0 and 8.0, respectively. In the case of the C-lobe of PO system, all the iron coordinating residues get more distant from $Fe^{+3}$ due to triggering by tyrosine protonation, resulting in opening of the C-lobe that is originally closed in PO. Most of the change that leads to the 5.1±0.4 Å RMSD in the PO system (figure 2b) is due to the mobility of the linker region. Thus, within 200 ns, the H structure evolves to a conformation that is also sampled by the equilibrated PO structure and we conclude that the experimentally observed PO state is a stable intermediate of hTf once N-lobe opening is triggered.

Finally, we study the dynamics of apo hTf (figure 2c). We find that, in the absence of irons, the N-lobe is now more restricted in its motions compared to the C-lobe. Moreover, the overall RMSD of the protein displays large deviations from the initial structure, exceeding 5.5±0.9 Å. However, these fluctuations do not signify sampling of alternative microstates of the molecule. Rather, in the absence of the locking effect of the irons, the motions of the protein, especially between the N- and the C-lobe, and by the hinge between $C_1$-$C_2$ subdomains are dominant. To quantify this observation, we calculate the RMSD between the average structures of the four consecutive 50 ns portions from the A run. The values are as small as 1.4 Å (between the average structures in the 100-150 and 150-200 ns portions) and as large as 2.8 Å (between the 50-100 and 150-200 ns portions). Note that these values are on par with the resolution of the crystal structure of the apo form (2.7 Å). We also monitor the eigenvectors belonging to the slowest mode of motion, obtained from the principle component analysis of the 50 ns subsets of the 200 ns long trajectory. We find that the overlap between the dominant eigenvector of each chunk is larger than 0.5 in all cases, for the overall protein, and for only the C-lobe. As a sample, the dominant eigenvector of the C-lobe is displayed in supplementary figure S4 for the four 50 ns trajectory pieces, signifying the same hinge motion around the iron binding region in the absence of the iron. Thus, despite the large RMSD recorded during the MD simulations, the apo form is a loose structure displaying large fluctuations around its experimentally determined state and does not make a conformational jump to another microstate.

According to the RMSD curves, we hypothesize that at serum pH, upon triggering lobe opening, the conformation travels through the partially open form before adopting the open conformation. To further scrutinize the details of this conformational change, we display in figure 3, the locations of conformational space sampled during the MD simulation of different systems projected on the reduced degrees of freedom (defined in figure 1; $\theta_N$, $\theta_C$, $\varphi$). In the figure, regions sampled by $\theta_N$-$\theta_C$ are displayed on the leftmost column as contour maps to show the more populated areas of these geometric parameters. The position of the initial x-ray structures are marked by the red dots. The center column displays the distribution of the relative positioning of the N- and the C-lobes, projected on the angle $\varphi$. Also displayed on the right side of the figure is the superposition of the full structure snapshots of corresponding systems sampled during the 200 ns at 4 ns intervals.

Comparing the distributions of the reduced degrees of freedom for the H and PO systems (figure 3), we find that the N-lobe of both systems samples bending angles in the range $\theta_N = 120\pm10°$. Interestingly, the C-lobe of the PO samples a wider range than the N-lobe, although this behavior is not directly evident from the RMSD curves in figure 2. The most populated region by both is nevertheless the same, centered near $\theta_C = 100°$. The relative orientation of the N- and the C-lobes for these systems, quantified by $\varphi$ are nearly the same by the end of the simulations. The second hump in the $\varphi$ distribution of H appears later in the simulations, once the conformational change in the C-lobe is also triggered at around 100 ns time point. The final distribution of $\varphi$ is centered on $-20\pm10°$ in both H and PO systems. Thus, there is substantial overlap between the regions populated the N-lobe of H and PO systems, corroborating our previous observation that the PO state is a stable intermediate of the H state at serum pH.

The above-mentioned results provide a good description of how the H system is driven towards the regions sampled by the PO conformer, once the change is triggered by the initial conditions described earlier. Furthermore, comparison of the $\theta_N$-$\theta_C$ maps and $\varphi$ distributions of the H and A (figure 3) portrays the fact that, in spite of the opening motions of the N-lobe and large mobility of the C-lobe, H has more propensity toward PO than A. The bending angle in the C-lobe of A, $\theta_C$, overlaps significantly with that of either H and PO, while in the N-lobe, $\theta_N$ displays a small but discernible shift by 10° centered on the range $\theta_N = 110\pm10°$. More significantly, the relative positioning of the two lobes, quantified by $\varphi$, is displaced by ca. 50°, centered on ca. $-70\pm30°$ in A. Thus, the switch to the apo structure requires the large motion of the two lobes sliding past each other, besides the opening of the $Fe^{+3}$ binding site. This might require much longer time scale events than can be observed by the current conditions of the MD simulations.

**Dilysine trigger is not strong enough to destabilize the $Fe^{+3}$ binding cleft and induce domain opening.** It has been proposed that either K206 or K296 are deprotonated in holo hTf and therefore are able to tightly interact via a hydrogen bond to lock the $N_1$-$N_2$ subdomains [14]. Therein, it was suggested that the protonation of these lysines would lead to strong enough repulsions in the second shell to force open the iron binding cleft. These residues were therefore labeled "dilysine trigger." On the other hand, previous MD simulations of N-lobe hTf suggested that protonation of these lysines was not the sole reason of domain opening [11]. However, therein the simulations were carried out on closed form of the apo protein (i.e. iron was not included in the simulations) and was of length 1.5 ns. Our MD simulations provide a testbed for this discussion since we take standard protonation state of these two lysines (i.e. they are charged and are expected to repel each other; therefore, they might trigger a loosened iron binding cleft).

In H, the two lysines are initially 4.0 Å apart and their separation (measured from the ε-amino group) grows up to 19.0 Å, along with N-lobe opening. If a strong repulsion between them were at the route of the trigger, similar observations should be made for the H* and H** systems which have the same Lys protonation states. However, our observation is contrary to this assumption: in H* the dilysine distance changes from 4.4 Å up to 6.5 Å and in H** from 3.9 Å to 8.4 Å. In neither of these runs does the N-lobe open (see Table 3 for the final iron coordination distances in each of these system). Thus, we find that opening of the N-lobe steered by the protonation states of the tyrosine residues is the main reason for the two lysines to grow apart and open the cleft.

This observation emphasizes the importance of protonation states of the two liganding tyrosines and the fact that dilysine trigger, even when the $Fe^{+3}$ is loosely connected as in the H* run, does not play a key role in opening the iron binding cleft. This is reasonable in light of the fact that in bovine lactoferrin structure, the two lysines equivalent to those in hTf dilysine trigger do not participate in a hydrogen-bond network and yet the protein is capable of undergoing the conformational changes necessary for iron release [42]. In the same manner, homologous residues (K534 and R641) in pig Tf are too distant to share a hydrogen bond [43]. We therefore turn our attention to the first shell residues.

The unusually low $pK_a$ of Y188 (see Table S1 and ref. [11]) suggests that it might be charged in the holo form and therefore might resist losing contact with $Fe^{+3}$. Kinetic data also provides evidence for contributions from tyrosine residues and mainly report the conformational changes which precede and allow iron release [44]. Site-directed mutagenesis of Asp and His residues in the binding region changes the binding properties of the protein, but the protein never loses its ability to bind iron even when the coordinating residue types have been mutated to noncoordinating ones such as Ala or Phe. Conversely, the Y188F mutation completely abolishes the natural ferric complex with carbonate and iron binding ability [39]. On the other hand, Y517F mutant leads to a complete absence of iron binding in the C-lobe, while Y426F mutant has extremely weak binding properties [45].

**At serum pH, fluctuations of the N-lobe are communicated to the C-lobe through the peptide linker**. We next investigate if there is crosstalk between the two lobes. We therefore record the root mean square fluctuations (RMSF) of the H system in 50 ns time intervals. We display in figure 4a the time evolution of fluctuations resulting in global rearrangements of the two lobes in the full structure. The regions with large RMSF are all confined to unstructured loops. In the first 50 ns MD simulation of the H system, large fluctuations mostly occur in the N-lobe. In the time interval of 50-100 ns, the unstructured linker undergoes larger fluctuations, the effect of which is later communicated to the C-lobe. After 150 ns of MD simulation, we observe propagated dynamics throughout the full structure. In particular, fluctuations in residues 488-519 (shaded in gray) are accentuated over time becoming the most dominant in the C-lobe as well as the whole structure, while those in the N-lobe residues 159-189 and 274-295 (shaded in blue) are diminished. These residues are colored in supporting figure S1. They control the exit pathways of the $Fe^{+3}$ ion in the lobes (shown by the arrows in figure S1). Since previous MD simulations on hTf involved only the holo N-lobe structure in the absence of iron, this is the first work where such cooperativity has been observed.

Accordingly, we compare the RMSD trajectories of the two lobes in the full structure to those from 200 ns MD simulations of isolated lobes of hTf, labeled as $H_N$ and $H_C$ (supporting figure S5). We note that the lobe opening is "triggered" in these simulations as well by neutralizing the $Fe^{+3}$ coordinating tyrosine residues. The RMSD profile of the $H_N$ is similar to that of the N-lobe in the full structure of H, where both of them advance towards an open conformer starting within the first 20 ns of the simulation. Conversely, based on the RMSD of $H_C$, we observe a more compact and stable C-lobe in the absence of N-lobe compared to its counterpart in the full structure. The RMSF profiles of these systems are provided in figure 4b. We find that the relative changes in the peak positions of the fluctuations in loops 159-189 and 274-295 also occur in the isolated system $H_N$. Thus, these conformational dynamics are intrinsic to the N-lobe. On the other hand, the large fluctuations in the 488-519 loop of the C-lobe observed after 100 ns in the full structure (figure 4a gray shaded region) are absent throughout the $H_C$ system simulations (figure 4b gray shaded region), implying that they are triggered by the mobility in the N-lobe in the full structure. Thus, the C-lobe has more intrinsic stability than the N-lobe, and the latter acts as a pulling machine that stimulates the C-lobe to open in the full hTf structure.

As a final word on the isolated lobe simulations, we note that while the RMSD change is low in the C-lobe, the iron coordination is nevertheless disrupted. In fact, the iron coordination distances observed by 200 ns for the N-lobe only and C-lobe only systems ($H_N$ and $H_C$) are similar to those from the H system (Table 3). That the overall structure remains almost intact while the iron coordination is lost points to the fact that the overall structure of the C-lobe contributes to stabilizing the conformations.

**At serum pH, N-lobe is the iron carrier; C-lobe provides the signal for external recognition.** In the H system, we take the two liganding tyrosines in each lobe in their protonated state. We correlate the opening fluctuations of the N-lobe to the absence of strong electrostatic interactions between the tyrosines and iron. Therefore, the N-lobe is free to undergo rigid body motions and open the iron-containing cleft. We note that the C-lobe does not respond to the triggering in the same manner, despite having the same modification in the charge state of the liganding tyrosines. The conformational change prior to iron release is a crucial initial step in restricting/accelerating this process. Kinetic experiments also indicate the rate of iron release is slower in the C-lobe (0.65±0.06 $min^{-1}$) compared to the N-lobe (17.7±2.2 $min^{-1}$) in hTf [37,46]. Presence of a highly conserved extra triplet of disulphide bridges in the C-lobe as compared to the N-lobe presents one explanation for the slower iron release rate in the C-lobe [15].

One of the disulphide bridges on the C-lobe is located in the interdomain cleft where the protein tightly binds iron and restricts domain movement. Presence of this disulphide bridge has no effect on iron release and uptake properties, but significantly increases the conformational stability and prevents the conformation of the C-lobe from becoming loose or open [15]. Furthermore, the hinge region in the C-lobe (residue 424-433) is longer than that in the N-lobe (residue 94-100), while the C-lobe as a whole is more rigid than the N-lobe due to the presence of additional disulphide bridges. The control of the binding cleft in the latter is therefore accomplished by the internal linker connecting the rigidified $C_1$ and $C_2$ subdomains [44]. In addition, the C-lobe of hTf is shown to play a major role of hTf recognition by other proteins. A two step mechanism proposed for hTf-TfR complex formation involves a fast (50 μs) and very strong interaction between the C-lobe of hTf and TfR, accompanied by a slow (2-3 hr) and very weak interaction of TfR and N-lobe of hTf [47]. In particular, the C-lobe serves as a TfR recognition region in addition to being an iron transporter; conversely, the N-lobe functions only as an iron carrier and therefore exhibits more labile and flexible tertiary structure [6,41]. Therefore, the lag of the C-lobe opening behind that of the N-lobe assigns them the respective main roles of recognition lobe and iron carrier.

The interaction of hTf with its receptor TfR and internalization during endocytosis is efficient for iron-bound forms while it is absent for apo hTf [48,49]. This interplay of hTf-TfR affinity is thought to play a critical role in recycling apo hTf back into the serum, but how this affinity change actually occurs has been raised as an important question [48]. The stability of the C-lobe over the N-lobe at serum pH is also corroborated by our 200 ns MD simulations on monoferric forms of the full structure ($H^{Fe+C}$ and $H^{Fe+N}$ systems in Table 2). Comparison of overall RMSD of H, $H^{Fe+N}$ and $H^{Fe+C}$ systems within the first 50 ns specifies that upon triggering opening, the N-lobe always has higher mobility than the C-lobe. Moreover, the diferric form has the highest overall mobility while $H^{Fe+N}$ has the lowest (figure S6). We note that the monoferric form with iron in the N-lobe is twice as likely to be found in monoferric hTf pool than that with iron in the C-lobe [50].

In the monoferric hTf simulations we make two interesting observations: (i) The N-lobe is more mobile than the C-lobe even in monoferric hTf with iron in the N-lobe, similar to the diferric form. (ii) In monoferric hTf with iron in the C-lobe, after ca. 80 ns, the C-lobe gets to be more mobile than the N-lobe despite the associating role of iron present in the former. Thus, once iron dissociates from the N-lobe in diferric form (e.g. in a system similar to H), the C-lobe gains additional mobility and has higher propensity to dissociate the second iron. In fact, in the $H^{Fe+C}$ system, we find that the coordinating Asp is not as tightly bound to the iron as in the rest of the systems except for the PO structure which is the most similar system to the $H^{Fe+C}$ in terms of structure (coordination distance increases from 1.9 Å to 4.2 Å; shown in bold in Table 3).

In sum, for both monoferric and diferric hTf, the C-lobe mostly displays lower mobility than the N-lobe. The situation is reversed for apo hTf, making the C-lobe too mobile for lock-and-key recognition by the receptor. The interplay between the flexibilities of the two lobes in the presence/absence of the irons makes the two-lobed hTf optimized for both its recognition on the cell surface and then its release into the serum, leading to a perfect recycling system for iron transport.

**Endosomal pH prompts hTf to an open conformation.** The mildy acidic pH of the endosome is the major factor effecting iron release during receptor-mediated endocytosis [12]. We mimic the endosomal environment (pH~5.6) in $H^{\dagger}$ system by changing the protonation of the $pK_a$ shifted residues mentioned under the Models and Methods section. We observe some specific structural changes that are different from all other systems studied in this work. The most prominent effect is in the simultaneous opening of the two lobes, as opposed to the delayed opening of the C-lobe observed at serum pH (compare figure 2a and 2d). Thus, at low pH, hTf proceeds from the closed conformation to the open form directly, without the requirement of passing through the PO structure.

In the H$^\dagger$ system, Y95 and Y188 in the N-lobe move away from the iron, reaching distances of ca. 10 Å, even further than that observed in the H system. In the C-lobe, Y517 is more distant than Y426. Coordinating Asp and His residues never leave the active site, preventing the two subdomains in each lobe to reach the completely open conformation.

The sequence of events in H$^\dagger$ is examined in more detail through the distributions of the reduced degrees of freedom $\theta_N$, $\theta_C$, $\varphi$, displayed in figure 3, bottom row. $\theta_N$ and $\theta_C$ mostly sample the exact same region as A up to 130 ns. Thereafter, the $\theta_C$ angle equilibrates at the region around 125°. This is extreme opening of the C-lobe, also observed in the PO structure to some extent, exposing the iron to chelators for proper dissociation. On the other hand, the $\varphi$ distribution shifts to the region centered around -70° between 115 - 125 ns, also sampled by A. It then goes back to the original $\varphi$ angle region near 0° at 130 ns. Thus, once the pH is lowered, the two lobes are able to make the large motion of moving past each other by ca. 50° multiple times, without shedding the Fe$^{+3}$ ions. The lower pH has a lubricating effect on the conformations of hTf. The above-mentioned results provide us with an excellent description of how H$^\dagger$ system is driven directly toward open conformer under endosomal pH within the observation window of 200 ns MD simulations.

# Conclusion

We have gained a broad view into the dynamical properties of hTf by carrying out extensive MD simulations on three different experimentally resolved conformations of hTf (H, PO, A), its two monoferric forms ($H^{Fe+N}$, $H^{Fe+C}$) and the two separate lobes in iron bound conformation ($H_N$, $H_C$). The simulations have implications for iron-binding/release mechanisms of hTf, as well as its recognition at the cell surface in the iron loaded form and release by the cell in the iron free form. Our findings provide a first step towards our long term goal to simulate an in vivo environment that adequately models the functioning of transferrin.

In the PO structure, the N-lobe is similar to that of the apo form (1.7 Å RMSD) and the C-lobe is similar to that of the holo form (0.5 Å RMSD); Table 1. We show that the holo conformation may be destabilized by manipulating protonation states of some distinct residues (figure S2 a-c). Large conformational changes may be delimited or induced by targeting perturbation-sensitive residues selected based-on their predicted $pK_a$ values (figure 2a). By mimicking two different pH environments and monitoring reduced degrees of freedom for intra and inter-lobe motions, we distinguish two possible pathways for conformational changes occurring prior to iron release. At serum pH, we observe the closed → PO transformation within 200 ns as a result of charging two tyrosine residues in the iron-binding cleft; this structure remains stable up to 0.5 µs. We expect PO → open transformation to follow, but on a longer time scale, since the two lobes cannot make the necessary step of moving past each other by $\varphi = 50°$ under these conditions (figure 3, center column). Yet, the distance of iron to its coordinating residues within 200 ns in the PO structure implies that the dissociation event towards the A conformation has already begun (Table 3). Conversely, at the lower pH of the endosome, a direct closed → open switch is observed within 200 ns (figure 2d); the two lobes sliding past each other is thus triggered by lowering the pH.

Essential dynamics analysis show that the open form experiences hinge-bending and hinge-twist as its main modes of motion of the subdomains and domains, respectively [11,13]. Accordingly, these motions are proposed to facilitate iron binding, since they help the open form to sample conformations between that of crystallographic open and closed forms. We find that monitoring the RMSD trajectories provides an incomplete picture of the conformations sampled by hTf. While the A system displays large changes compared to its x-ray structure of up to 8 Å (figure 2c), it actually samples a single wide microstate since the structures averaged over 50 ns subsections are the same, as are the essential modes sampled in each 50 ns time window (figure S4). Conversely, the $H^{\dagger}$ system goes through different microstates, experiencing reversible gliding of the two lobes past each other multiple times, a motion not observed at serum pH figure 3, center column).

Consequently, we raise the question on the role of dilysine trigger in the opening behavior of the N-lobe. We make a set of knock-out designs of simulations where the two lysines are charged but the iron coordinating tyrosines are protonated to provide three different levels of affinity to the ion in holo hTf (H, H* and H**). We find, by comparing the H* and H** with the H and $H_N$ systems (Table 3, figures S2, S5 and Table S2), that the repulsion between the two lysines is not large enough to be the sole reason for triggering subdomain separation.

In systems where domain separation occurs (H, PO, $H^{Fe+N}$, $H^{Fe+C}$, $H_N$, $H^{\dagger}$), iron is observed to progress toward the outside of the binding cleft, but never gets released as a result strong interactions with synergistic anion and inhibited by the absence of other environmental factors such as chelators. We note that in the absence of suitable iron chelators and TfR, mild acidification of the endosome is not sufficient to stimulate iron disassociation from the cleft [32,37,51]. In both lobes, carbonate anion does not move away from its original position in the binding site as also evidenced by experiments where it remains even after iron is released from both lobes [22].

At endosomal pH, we observe a simultaneous conformational switch of the two lobes as opposed to the delayed switch at serum pH. The observed behavior in the MD simulations on individual lobes indicates that the motions of the two are concerted so as to facilitate the controlled release of iron at serum pH. The N-lobe acts as a pulling machine that stimulates the C-lobe to open in the full hTf structure. Therefore, the bilobal nature of hTf is a biological advantage, promoting iron release as a result of interlobe communication.

The above discussion refers to the iron dissociation from hTf under different conditions. On a larger length scale, our results also hint at how hTf as a whole is captured and released at the cell surface. By monitoring the relative mobility of the N- and the C-lobes, we find that in iron bound form, the C-lobe has more intrinsic stability than the N-lobe, while the situation is reversed in iron free hTf. The C-lobe of iron loaded hTf provides both the specific recognition sites on its surface and a stable template for TfR recognition at serum pH. In iron free form, although the surface sites are still available for TfR binding, the fluctuations are too high for holding them together, facilitating easy discharge at the cell surface.

We conjecture that the pathogens may also be utilizing the same mechanism as TfR for iron sequestration from hTf. Micro-organisms such as *Haemophilus influenzae, Neisseria meningitides* and *Neisseria gonorrhoeae* have developed efficient mechanisms for stealing iron directly from the plasma [52]. Iron discharge by a gram-negative bacteria, *Neisseria,* through its transport machinery (TbpA and TbpB) is also preceded by selectively binding hTf C-lobe at physiological pH to mediate direct iron extraction from the human host [10]. Extracellular compartment of the TbpA solely interacts via an unusually long plug loop with $C_1$ subdomain of the hTf. Simultaneously, an extracellular loop is inserted into the cleft between $C_1$ and $C_2$ subdomain. Disruption of the second shell residues facilitates iron import to the chamber formed between TbpA and TbpA β-barrel [53]. The conformational change induced under these circumstances is a partial opening of the C-lobe to that between apo and holo forms and disturbance of the iron coordination. Since *Neisseria* is not able to consume both irons stored in hTf, we conclude that the compact and stable tertiary structure of the C-lobe at physiological pH may have been developed as a defense strategy to prevent the iron acquisition by pathogens.

# Declaration of interest

The authors report no conflicts of interest. The authors alone are responsible for the content and writing of the paper.

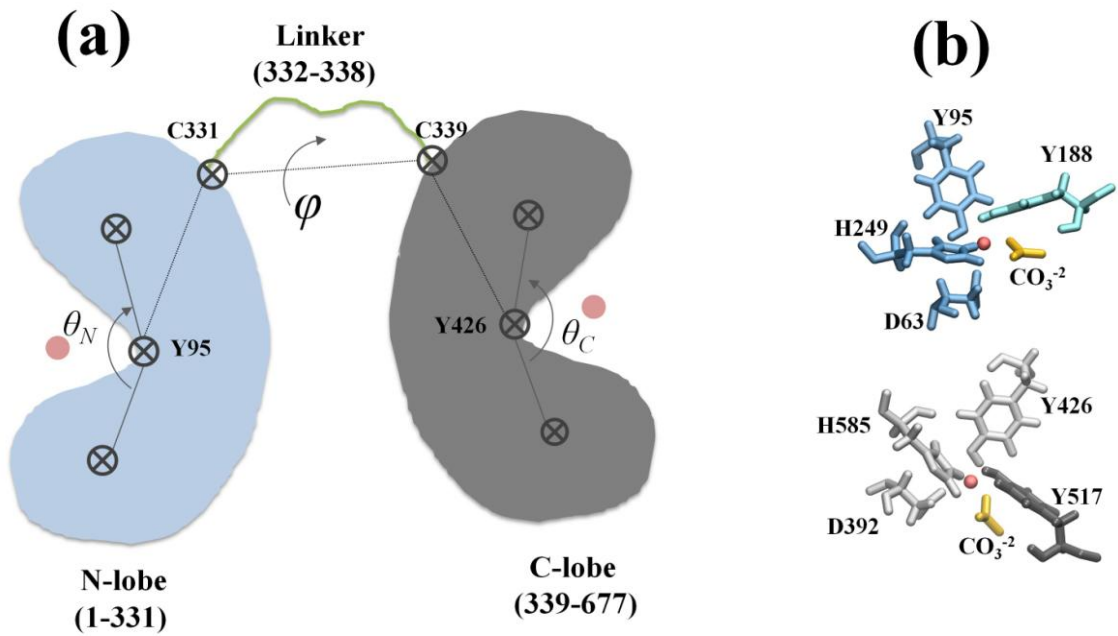

Figure 1

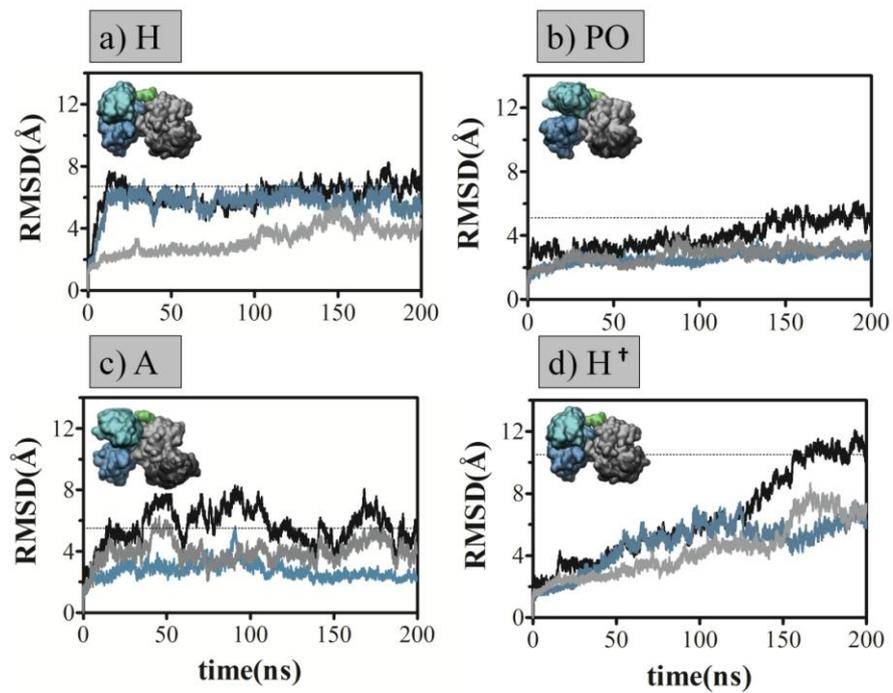

Figure 2

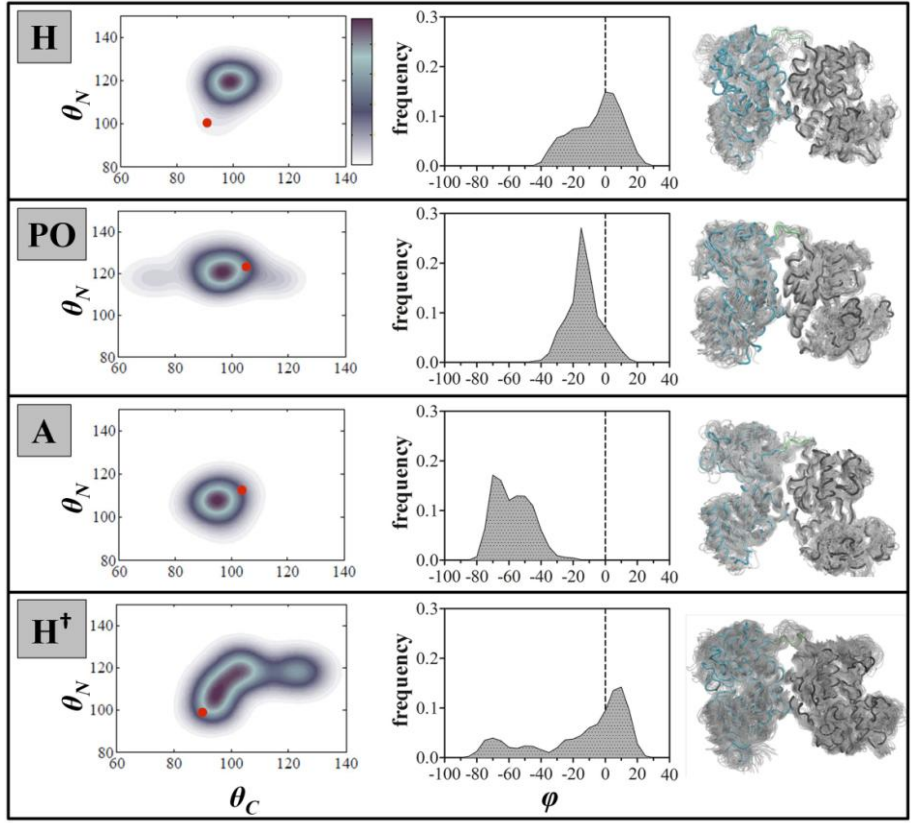

Figure 3

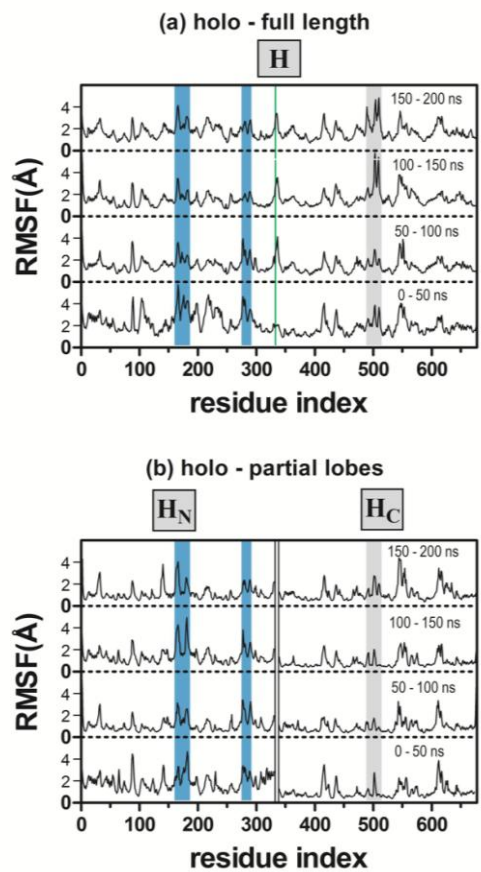

Figure 4

# Figure Legends

**Figure 1.** (a) Schematics of hTf displaying the color code used throughout the text (blue for N-lobe, green for linker, gray for C-lobe; pink circles represent $Fe^{+3}$ ions), the points used to define the reduced degrees of freedom (shown by $\otimes$) are the centers of mass of the N- and C-lobes, the beginning and end points of the linker, and the hinge residues between the subdomains of each domain, Y95 and Y426. We define the lobe bending angles $\theta_N$ and $\theta_C$, measured by the angle between the two centers of mass and the hinge residue between each subdomain of the respective N- and C-lobes. The lobe torsional angle $\varphi$, is defined by the four points shown. These are the major degrees of freedom on which we project all trajectories to monitor the sampled conformations. (b) Iron (pink) coordinating residues in the N- and C-lobes. Residues in blue coordinate iron from the $N_1$ subdomain and residue in cyan is the sole coordinating residue from the $N_2$ subdomain. Residues in silver coordinate iron from the $C_1$ subdomain and residue in gray is the sole coordinating residue from the $C_2$ subdomain.

**Figure 2.** RMSD curves for H, PO and A systems (respectively, holo, partially-open and apo hTf at physiological pH) and $H^{\dagger}$ system (holo hTf at endosomal pH). The overall RMSD (black) as well as that of the N- and the C-lobes (blue and gray, respectively) are reported. Insets display surface representation of crystal structure for each system. Horizontal dotted lines represent the overall RMSD averaged over last 50 ns of each simulation. (a) In H, N-lobe displays large rearrangements from the beginning of the MD simulation. After 100 ns, these are communicated to the C-lobe. (b) In PO, both the N- and C-lobe stays near the initial conformer. The overall structure displays larger RMSD values due to relative placement of the lobes, communicated by the linker. (c) In A, the C-lobe is more mobile than the N-lobe throughout. It oscillates around the equilibrium state in a hinge like motion (see text for details). (d) In $H^{\dagger}$ where the lower pH of the endosome is mimicked, large rearrangements occur simultaneously in both lobes.

**Figure 3.** Conformations sampled by H, PO, A, $H^{\dagger}$ displayed by projecting on the reduced degrees of freedom, defined in figure 1a. The first column shows the joint probability distribution of the $\theta_N, \theta_C$ intra-lobe bending angles, therefore the differences in their sampled regions. The initial position is shown by the red dot: While the inequality is expected for the PO structure in the x-ray structure, even

the holo and the apo forms are positioned at slightly different values; the N-lobe is slightly more open in both cases. The second column shows the distribution of the torsional angle ($\varphi$) quantifying the relative positioning of the two lobes. Initially they are at -15, -30, -40, and -5°, respectively for H, PO, A, H†. The relative position of the iron loaded forms at serum pH is ca. 50° different than the iron free form. The endosomal pH system is more labile and samples both positions frequented by the apo and holo systems at higher pH. Shown on the right are the superposed structures of the corresponding systems, 50 snapshots spaced at 4 ns intervals.

**Figure 4.** $C_\alpha$ root mean square fluctuations averaged over equally spaced trajectory pieces of 50 ns each. a) RMSF of H system in black, b) RMSF of $H_N$ system is on the left (residues 1-331) and $H_C$ on the right (residues 339-677). Regions with more pronounced fluctuations in the N-lobe, linker and the C-lobe are highlighted in blue, green and gray, respectively. In the case of H system, in the first 50 ns, more pronounced fluctuations belong to the N-lobe. After 50 ns, the linker initiates fluctuations and the delayed communication between the lobes prompts the C-lobe to go through larger fluctuations after 100 ns. For holo hTf at pH=5.6, large fluctuations begin from the starting point of simulation.

# Tables

**Table 1.** RMSD between x-ray structure pairs*

|  | holo (3v83) | apo (2hav) | PO (3qyt) |
|---|---|---|---|
| holo (3V83) | <u>3.2</u> | **6.4**, *6.6* | **5.2**, *0.5* |
| apo (2HAV) | 8.2 | <u>3.4</u> | **1.7**, *6.6* |
| PO (3QYT) | 4.9 | 6.3 | <u>6.1</u> |

\* Upper diagonal, RMSD between N lobes (**Bold**) and C lobes (*Italic*); Diagonal (underlined), RMSD between N lobe and C lobe within each structure; lower diagonal, overall RMSD between every pair.

**Table 2.** Summary of system conditions for different simulations.

| System label | Perturbation scenario | Number of atoms ($Na^+$, $Cl^-$) | Equilibrated box size ($Å^3$) | Simulation length (ns) |
|---|---|---|---|---|
| H** | All tyrosines in contact with $Fe^{+3}$ deprotonated on both lobes (locked irons) | 78214 (70,63) | 88×107×90 | 60 |
| H* | One tyrosine contacting $Fe^{+3}$ from each lobe deprotonated (Y188 and Y517) | 77947 (68,63) | 88×107×90 | 60 |
| H | All residues in standard protonation states | 78214 (66,63) | 88×107×90 | 500,150 |
| A | All residues in standard protonation states | 86426 (73,71) | 97×99×96 | 200 |
| PO | All residues in standard protonation states | 79270 (66,65) | 88×105×91 | 200 |
| $H^{Fe+N}$ | $Fe^{+3}$ only in N-lobe | 77940 (68,63) | 88×107×90 | 200 |
| $H^{Fe+C}$ | $Fe^{+3}$ only in C-lobe | 77968 (68,63) | 88×107×90 | 200 |
| $H_N$ | N-lobe only; $Fe^{+3}$ included | 34781 (28,28) | 85×60×73 | 200 |
| $H_C$ | C-lobe only; $Fe^{+3}$ included | 36957 (30,30) | 78×75×68 | 200 |
| $H^†$ | Histidines protonated as described under Models and Methods to mimic endosomal pH. | 77947 (63,69) | 88×107×90 | 200 |

**Table 3.** Distances from coordinating residues[1] to iron in the beginning[2]/end[3] of simulations.

| Liganding Residue | H** | H* | H | PO | $H^{Fe+N}$/$H^{Fe+C}$ | $H_N$/$H_C$ | $H^†$ |
|---|---|---|---|---|---|---|---|
| Y95 | 1.9/1.9 | 2.2/4.2 | 2.2/6.0 | 2.2/6.0 | 2.2/4.8 | 2.2/10.7 | 2.2/9.5 |
| Y188 | 1.9/2.0 | 1.9/1.9 | 2.3/6.4 | 2.2/8.0 | 2.8/5.9 | 2.8/10.6 | 4.0/10.8 |
| H249 | 1.9/3.7 | 3.8/3.7 | 2.2/4.5 | 10.1/4.3 | 4.2/5.2 | 4.2/5.5 | 4.3/4.2 |
| D63 | 3.4/2.1 | 1.9/2.0 | 1.9/2.1 | 6.0/2.1 | 2.0/2.0 | 1.9/2.0 | 2.0/2.1 |
| Y426 | 1.9/1.9 | 3.4/3.3 | 2.2/6.0 | 2.2/8.7 | 2.2/8.8 | 2.2/7.6 | 2.2/5.1 |
| Y517 | 1.9/2.0 | 1.9/2.1 | 2.2/6.1 | 2.3/9.0 | 2.2/6.5 | 2.2/8.3 | 2.2/6.9 |
| H585 | 3.3/4.3 | 3.8/4.8 | 1.9/2.3 | 2.2/5.4 | 4.9/4.2 | 4.9/5.9 | 4.6/4.1 |
| D392 | 1.9/2.0 | 1.9/2.0 | 2.2/2.1 | 1.9/**4.3** | 1.9/**4.2** | 2.0/2.0 | 2.0/2.1 |

[1] measured from carboxylate oxygen of D, imidazole nitrogen of H and phenolate oxygen of Y.
[2] obtained right after energy minimization of the system.
[3] that of the snapshot in the final 50 ns of the trajectory for which average distance between $Fe^{+3}$ and its four coordinating residues is minimum.

Figure S1.
Relative positioning of the subdomains and the schematics of the exit pathways of holo hTf

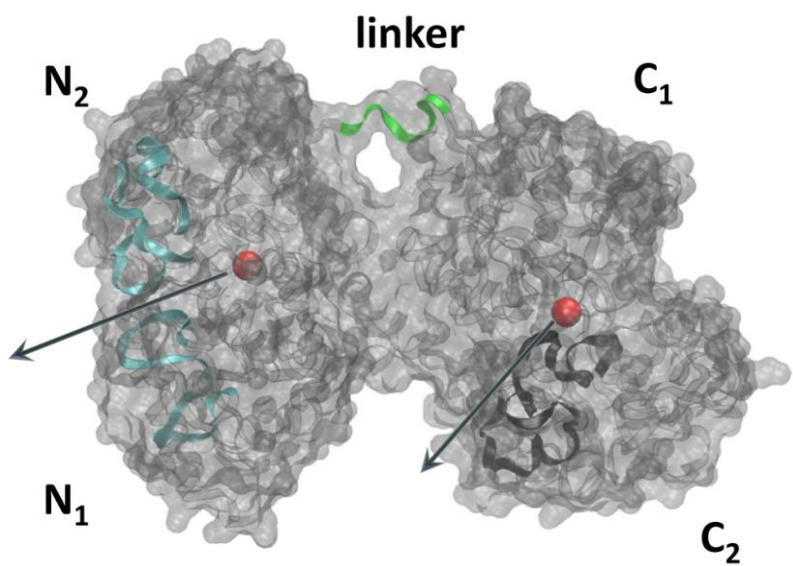

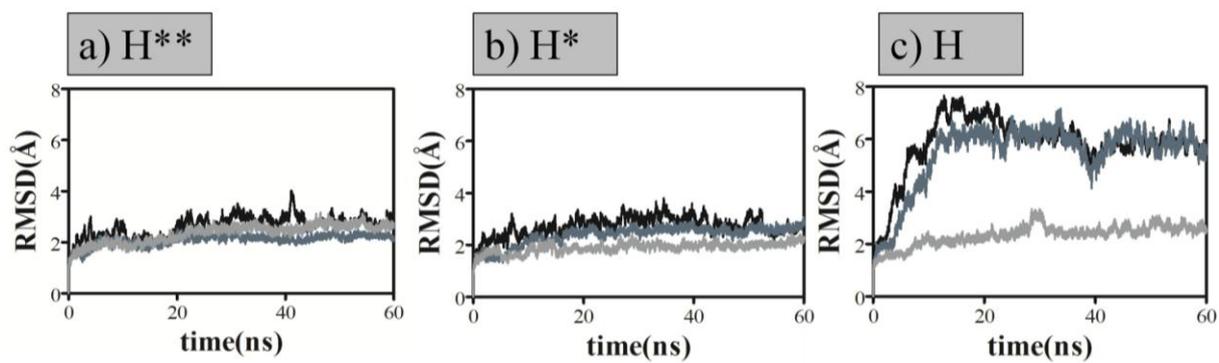

Figure S2

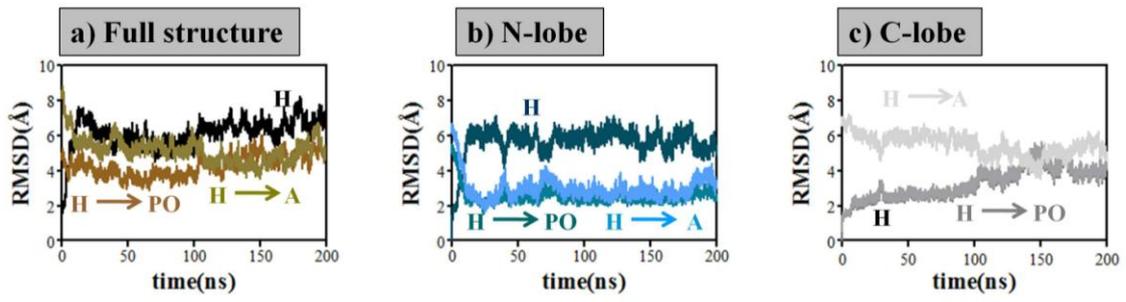

Figure S3

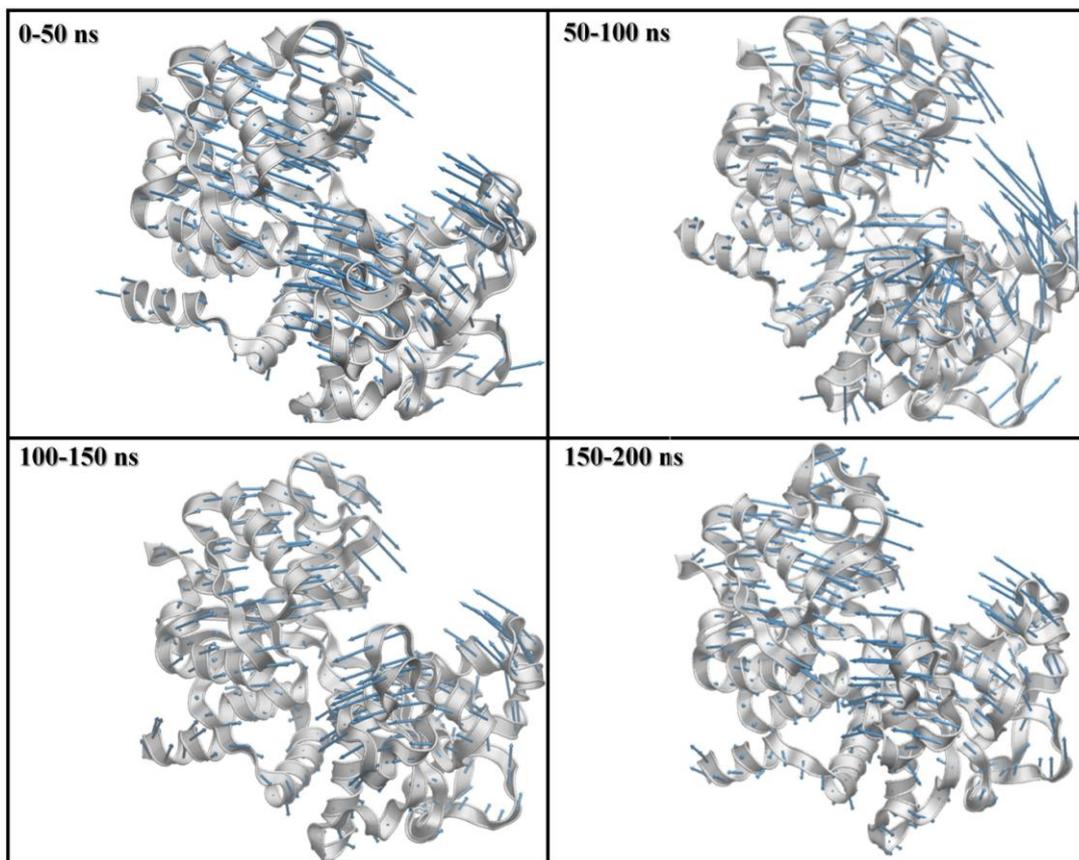

Figure S4

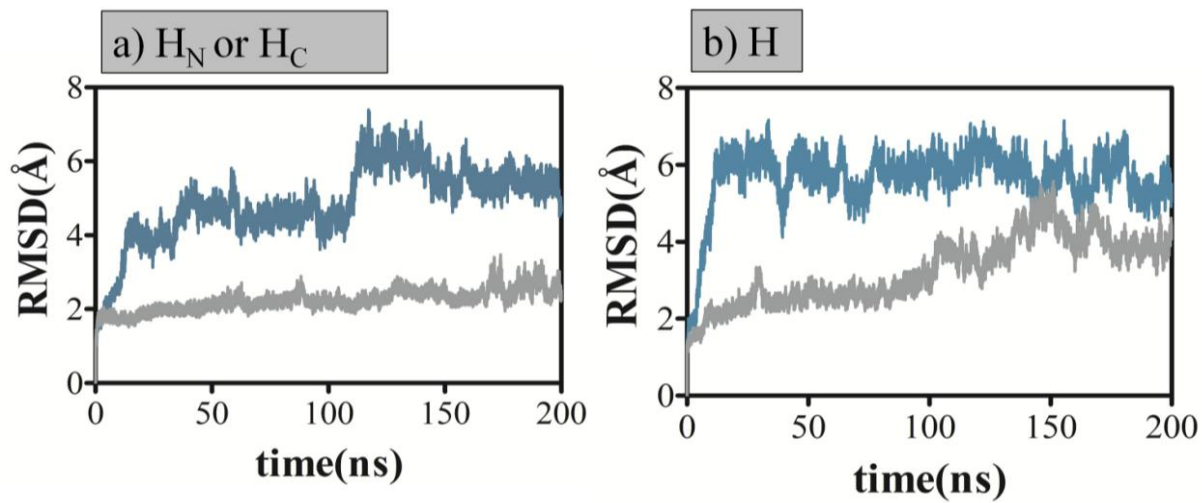

Figure S5

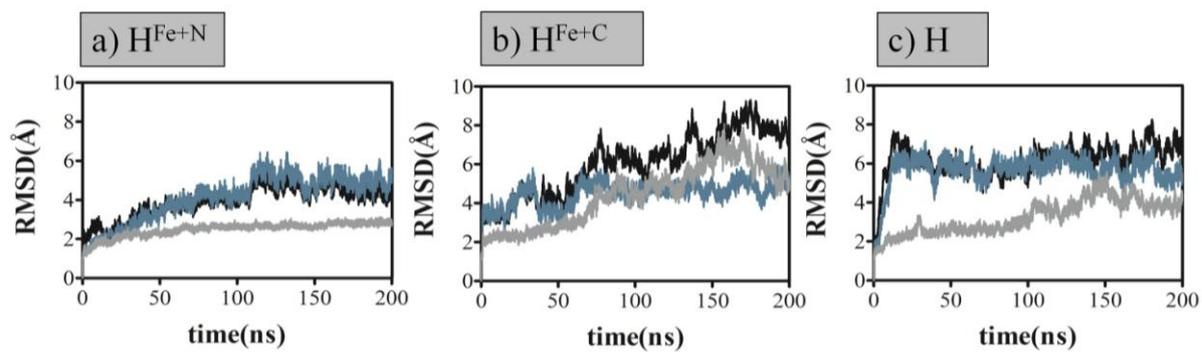

Figure S6

# Supporting Information Legends

**Figure S1.** Relative positioning of the subdomains and the schematics of the exit pathways of holo hTf. The two lobes are positioned such that the exit paths whose approximate directions are shown by the arrows are nearly orthogonal. The linker is shown in green. The loops that display large differences in the RMSF profiles are shown in blue on the N-lobe (residues 159-189 and 274-295) and dark gray in the C-lobe (residues 488-519); see figure 4 and the main text for details.

**Figure S2.** RMSD curves for H**, H* and H systems, labeled as defined in Table 2. The overall RMSD (black) as well as that of the N- and the C-lobes are reported (blue and gray, respectively). When both of the iron coordinating Tyr residues in the lobes (a) or only from each lobe (b) are negatively charged, the lobes are sealed tight. The N-lobe fluctuations are slightly more pronounced in (b) than in (a). The overall RMSD of the protein, as well as that of either lobe is below 3 Å throughout the time window of observation. (c) When all Tyr are in their standard protonation states, the N-lobe begins to open up within 10 ns while the C-lobe remains closed.

**Figure S3.** Time evolution of the RMSD of H system and its lobes from different crystal structures. (a) The original overall RMSD curve is shown in black (a replica of that in figure 2a), while the same trajectory is also superimposed on the PO and A structures (brown and green, respectively). Once the initial conformational change is completed within 10 ns, the newly attained conformation displays lower RMSD from the PO x-ray structure. (b) The original N-lobe RMSD curve is shown in dark blue (a replica of that in figure 2a), while the same trajectory is also superimposed on the PO and A structures (shown as lighter shades of blue). N-lobe of H system displays excellent resemblance to the N-lobe of partially open structure. (c) The original C-lobe RMSD curve is shown in dark gray (a replica of that in figure 2a), while the same trajectory is also superimposed on the PO and A structures (shown as lighter shades of gray).Since the RMSD between the x-ray structures for the C-lobe of H and PO structures is very low (0.6 Å, Table 1), the curve for H→PO comparison is nearly identical to the original curve. The difference from the x-ray structure of A diminishes throughout the trajectory.

**Figure S4.** Eigenvectors displaying the dominant mode of motion in the C-lobe of apo the structure, obtained from four 50 ns–long portions of the A simulation, superposed on the average structure of each portion. While the RMSD from the crystal structure gets as large as 8 Å in these simulations, the average structures are within 2.8 Å of each other. Simulations displays the same large hinge bending motions around the average structure as shown, and does not exhibit a conformational shift.

**Figure S5.** RMSD profiles of the (a) iron loaded isolated lobe $H_N$ and $H_C$ runs compared to those from (b) fully loaded H. The N-lobe is colored in blue, and the C-lobe in gray. The N-lobe has propensity of open in both the full structure and the single lobe case. The C-lobe, on the other hand, has an intrinsic stability that prevents it from opening unless triggered by the conformational changes in the N-lobe.

**Figure S6.** RMSD profiles of the (a,b) monoferric full length hTf, $H^{Fe+N}$ and $H^{Fe+C}$, runs compared to those from (c) fully loaded H. The N-lobe is colored in blue, and the C-lobe in gray. Originally, the N-lobe is more mobile than the C-lobe in all cases. In monoferric form with iron in the C-lobe, after 80 ns, the mobility of the C-lobe gets similar to that of the N-lobe despite the associating role of iron present in the former.

**Table S1.** pK$_a$ values[1] for holo hTf residues calculated by H++ and PROPKA servers.

| Residue | PROPKA[2,3] | H++ | Ref [11] | Residue | PROPKA[2,3] | H++ | Ref [11] | Residue | PROPKA[2,3] | H++ | Ref [11] |
|---|---|---|---|---|---|---|---|---|---|---|---|
| D58  | 1.1   | <0  | 0.0 | K18  | 8.1   | >12 | >12  | H119 | 3.9 | 1.5 | 5.0 |
| D63  | <0    | <0  | 0.0 | K206 | 7.9*  | 4.2 | 8.4  | H207 | 2.4 | 2.5 | 0.0 |
| D297 | 6.4*  | 2.1 | 3.8 | K296 | <0*   | >12 | >12  | H249 | <0  | <0  | 0.0 |
| D392 | <0    | <0  | -   | K354 | 7.8*  | >12 | -    | H300 | 4.8 | 5.6 | 2.6 |
| D633 | 6.8*  | <0  | -   | R124 | 8.9   | >12 | >12  | H350 | 4.9 | 2.3 | -   |
| D634 | <0    | <0  | -   |      |       |     |      | H451 | 3.7 | 1.2 | -   |
| E15  | 6.3   | <0  | 1.2 | Y95  | >12*  | >12 | 1.2  | H473 | 5.6 | 3.4 | -   |
| E212 | 7.6*  | 1.0 | 1.4 | Y188 | 6.0*  | 0.6 | <0.0 | H535 | 3.7 | 2.6 | -   |
| E351 | 6.5   | <0  | -   | Y426 | 10.6* | >12 | -    | H585 | <0  | <0  | -   |
| E375 | 5.5   | 0.2 | -   | Y517 | 4.1*  | <0  | -    | H598 | 5.2 | 3.7 | -   |
| E512 | 4.4   | <0  | -   |      |       |     |      |      |     |     |     |
| E573 | 4.8   | 2.1 | -   |      |       |     |      |      |     |     |     |

[1] Of the total number of 214 residues with ionizable groups, those with ΔpK$_a$ > 2 for D, E, K, R and those with pK$_a$ ≤ 5.6 for H are reported. Values in the pH window of 0 – 12 are shown.
[2] Standard pK$_a$ values of residues in PROPKA server are D: 3.8, E: 4.5, H: 6.5, Y: 10.0, K: 10.5, R: 12.5.
[3] Residues marked by * are coupled to others, and hence may display alternate pK$_a$ values.
[4] Reference [11] only reports residues from the N-lobe.

**Table S2.** Overall and regional RMSD (in Å) values for different protein groups averaged over the last 50 ns of the corresponding trajectory.

| Initial structure | Protein | N-Lobe | C-lobe |
|---|---|---|---|
| H*         | 2.8±0.3  | 2.5±0.2 | 1.9±0.1 |
| H**        | 2.7±0.3  | 2.1±0.1 | 2.4±0.3 |
| H          | 6.7±0.5  | 5.6±0.5 | 4.1±0.4 |
| A          | 5.5±0.9  | 2.4±0.2 | 4.3±0.6 |
| PO         | 5.1±0.4  | 2.8±0.2 | 3.2±0.2 |
| H$^{Fe+N}$ | 4.5±0.3  | 5.0±0.4 | 2.8±0.1 |
| H$^{Fe+C}$ | 7.9±0.5  | 4.9±0.5 | 6.1±0.7 |
| H$_N$      | ---      | 5.4±0.3 | ---     |
| H$_C$      | ---      | ---     | 2.5±0.2 |
| H$^†$      | 10.5±0.7 | 5.7±0.7 | 6.9±0.6 |